\documentclass[twocolumn,amssymb,fleqn]{revtex4} 
\usepackage{epsfig,amssymb,amsmath,graphicx,subfigure,hyperref  }  
\usepackage{color, setspace} 
\usepackage{multirow}
\usepackage{tabularx}

\newcommand{\be}{\begin{eqnarray}}   
\newcommand{\ee}{\end{eqnarray}}

\begin{document}

\title{Emergent dynamic structures and statistical law in spherical lattice gas
  automata}
\author{Zhenwei Yao} 
\affiliation{School of Physics and Astronomy, and Institute of Natural
Sciences, Shanghai Jiao Tong University, Shanghai 200240 China}
\begin{abstract}
Various lattice gas automata have been proposed in the past decades to simulate
physics and address a host of problems on collective dynamics arising in diverse fields.
In this work, we employ the lattice gas model defined on the sphere to
investigate the curvature driven dynamic structures and analyze the statistical
behaviors in equilibrium. Under the simple propagation and collision rules, we
show that the uniform collective movement of the particles on the sphere is
geometrically frustrated, leading to several non-equilibrium dynamic structures
not found in the planar lattice, such as the emergent bubble and vortex
structures. With the accumulation of the collision effect, the system ultimately
reaches equilibrium in the sense that the distribution of the coarse-grained
speed approaches the two-dimensional Maxwell-Boltzmann distribution despite the
population fluctuations in the coarse-grained cells. The emergent regularity in
the statistical behavior of the system is rationalized by mapping our system to
a generalized random walk model. This work demonstrates the capability of the
spherical lattice gas automaton in revealing the lattice-guided dynamic
structures and simulating the equilibrium physics. It suggests the promising
possibility of using lattice gas automata defined on various curved surfaces to
explore geometrically driven non-equilibrium physics. 
\end{abstract}
\maketitle

\section{Introduction}

Lattice gas models are powerful and flexible tools to address a host of fundamental
problems arising in equilibrium and non-equilibrium statistical
physics~\cite{lee1952statistical, baxter1982exactly, kadanoff1986two,
kadanoff1999order,bagnoli2001nature,bagnoli2005phase,preston2013modern}, hydrodynamics~\cite{hardy1973time, hardy1976molecular,
margolus1986cellular, Frisch1986, frisch1987lattice,
swift1996lattice,buick2000gravity, wolf2004lattice}, pattern
formation~\cite{wolfram1984cellular, wolfram1986theory,
chen1995lattice,wootton2001local, deutsch2007cellular,kapral2012chemical}, and dynamical
systems~\cite{koelman1990cellular,manneville2012cellular, kier2002cellular,kuurka2012topological}.
Tracking the evolution of many identical agents under a few prescribed
rules in computer programs, $\it {i.e.,}$ the idea of cellular
automata~\cite{von1966theory, wolfram1983statistical, vichniac1984simulating,preston2013modern},
represents an alternative to, rather than an approximation of, differential
equations in modeling the natural world~\cite{toffoli1984cellular,
vichniac1984simulating, cowburn2000room}. A major goal of physics is to build
the bridge between a microscopic (level 1) and a macroscopic (level 2)
description of reality~\cite{kadanoff1999order, feynman1982simulating}.  To this end,
various lattice gas automaton models have been proposed. At
level 1, the particles follow a few well-defined rules and hop over the lattice. And one obtains
level 2 phenomena by coarse graining concerned physical quantities like velocity and
particle number. The first fully deterministic lattice gas model, which was defined on
square lattice known as HPP model, was proposed by Hardy, de Pazzis, and Pomeau
to simulate flow phenomena~\cite{hardy1973time, hardy1976molecular}. And its
variant version FHP model for triangular lattice was later introduced by
Frisch, Hasslacher, and Pomeau~\cite{Frisch1986, frisch1987lattice}.  These
lattice gas models are able to simulate the Navier-Stokes equation in the
regime of incompressible fluid~\cite{Frisch1986, frisch1987lattice}, and
reproduce featured flow behaviors in real fluids like vortices and
wave propagation~\cite{kadanoff1986two, margolus1986cellular}. For the
simplicity of the rules and the capability of simulating physics, these lattice gas models
are reliable tools for establishing the connection between the two levels of
reality.

We generalize the FHP model to the spherical lattice, and explore the emergent dynamic
structures and equilibrium properties at level 2 arising from the microscopic
motion of many identical constituents. Here, the spherical lattice refers to a 
two-dimensional triangular
lattice that covers the entire sphere. We choose the spherical lattice based on
the following considerations. First, extensive studies in the community of
active matter physics have shown that curvature can create rich dynamic
structures not found in the planar
geometry~\cite{keber2014topology,marchetti2013hydrodynamics,yao2016dressed}.
The spherical lattice is naturally curved, and provides an ideal environment to
explore the lattice-guided non-equilibrium dynamic structures. In addition, due to
the compactness of the spherical geometry, the subtle influence of the
boundary condition can be avoided.  In our model, the particles hop over the
lattice according to a few well-defined collision and propagation rules. In short,
the particles propagate along the lattice unless two particles of opposite velocities
occupy the same vertex or three particles move into the same vertex whose total
velocity is zero; binary and triple collisions occur in these cases.

In this work, we first show that any orbit of a single particle over the
spherical lattice, regardless of its initial position and velocity, is closed
and has the identical period.  Consequently, a cluster of particles with
quasi-uniform initial velocity is observed to separate and merge as a whole
with the alternating convergence and divergence of the lattice. In addition to
the periodic global morphological transformations, we also discover featured
non-equilibrium dynamic modes, including deforming bubbles in their propagation,
  and localized vortex structures. The presence of the disclinations in the
  spherical lattice, in combination with the compactness of the spherical
  geometry, is responsible for all these emergent dynamic structures. The
  accumulation of the collision effect ultimately brings the system in
  equilibrium, which is signified by the convergence of the collision frequency
  curves.  Statistical analysis reveals that the number of particles in each
  coarse-grained cell is subject to the Gaussian type fluctuation. For modest
  population fluctuation in the coarse-grained cells, the coarse-grained speed
  still conforms to the two-dimensional Maxwell-Boltzmann distribution. These
  results are well rationalized by mapping our system to a generalized random
  walk model. We further discuss the collision-driven randomization of the
  coarse-grained particle velocity, by which the application of the random walk
  model to our system is well justified.  Simulations show that either binary or
  triple collisions between particles, even at a rate as low as $0.06\%$, is
  sufficient to produce the Maxwell-Boltzmann distribution. This work
  demonstrates that the spherical lattice gas automaton is capable of simulating
  the geometrically driven non-equilibrium structures and the geometry-independent
  equilibrium properties. Our model may be generalized to other curved surfaces
  as well as the almost-planar lattices with a designed mix of five- and seven-fold disclinations
  to explore the geometrically driven non-equilibrium physics.

\section{Model}

\begin{figure}[th]
\centering
\includegraphics[width=3.6in]{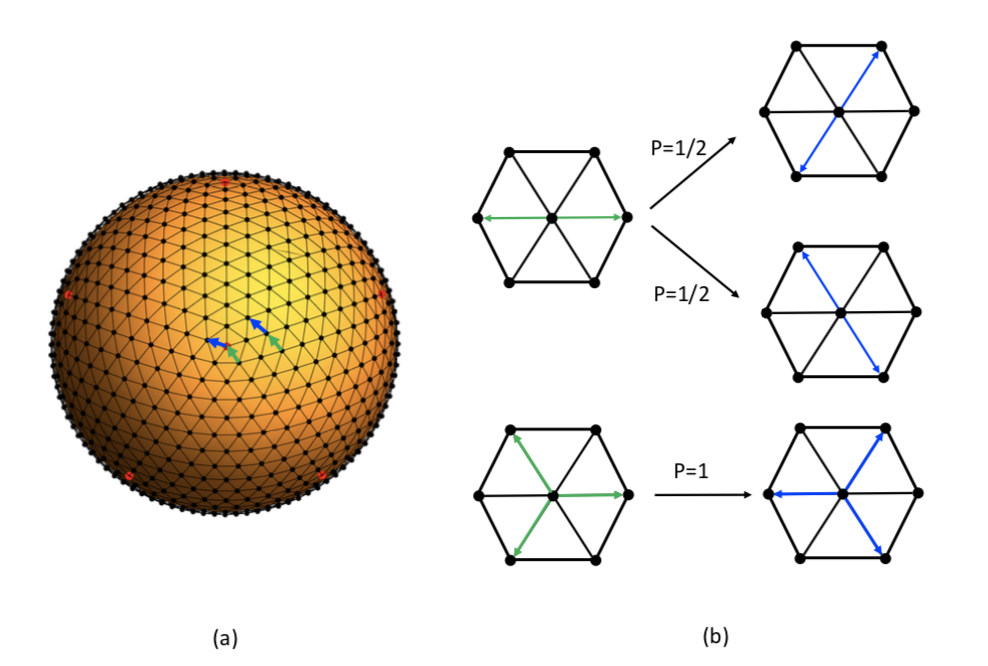}
\caption{Illustration of the generalized FHP lattice gas automaton model on the
sphere. (a) shows a spherical lattice where the red dots represent the five-fold
disclinations with five nearest neighbors. There are 12 five-fold disclinations
over the entire spherical lattice as a topological requirement. The green and blue arrows are to
explain the propagation rule on a regular and disclinational vertex.  (b) shows the
rules for the head-on binary and triple collisions.  
        } \label{model}
\end{figure}

We generalize the FHP model to the spherical lattice to study the particle dynamics
over the sphere.  The FHP model was originally designed on the planar triangular
lattice. By generalizing it to the spherical lattice, the model is modified
accordingly. In the spherical lattice, the point crystallographic defects called disclinations are
inevitable~\cite{chaikin2000principles}. An $n$-fold disclination in a triangular lattice refers to a vertex whose coordination
number $n$ is deviated from six. According to 
the Euler's
theorem~\cite{struik88a}, the total topological charge in a triangular lattice
is a topological invariant:
\begin{eqnarray}
    \sum_{i\in V}q_i = 6\chi,
\end{eqnarray}
where $q_i=6-z_i$ is the
topological charge of the vertex $i$, $z_i$ is the coordination number of the
vertex $i$, the sum is over all the vertices $V$ in the spherical lattice, and $\chi$ is the Euler's
characteristic. For the spherical topology,
$\chi=2$. It indicates that the total topological charge in the spherical lattice is 12.
The presence of the disclinations therein is therefore a topological requirement. 
The simplest spherical lattice contains uniformly distributed 12 five-fold
disclinations. We adopt such a spherical lattice in our work. 
The red dots in Fig.~\ref{model}(a) represent the five-fold disclinations. In the original FHP model, each vertex
accommodates up to six particles. In the spherical lattice, each five-fold
disclination can accommodate up to five particles.
The exclusion rule in the original FHP model is inherited in our model. That is, a vertex cannot accommodate two
or more particles of the same velocity.

In our model, the states of the particles are simultaneously updated by the following collision and propagation rules.
There are two types of collisions: the head-on binary and triple collisions.
The head-on binary collision occurs when a vertex is only occupied by two
particles of opposite velocities. In a triple collision, three particles meet at the same vertex whose
total velocity is zero.  The updated particle states after the two kinds of collisions are shown in
Fig.~\ref{model}(b). For the head-on binary collision, the two final states
occur with equal probability. The propagation rule is straightforward for a planar
triangular lattice: a particle moves along the lattice by the direction of its
velocity as shown in Fig.~\ref{model}(a). On a spherical lattice, all of the
particles must turn either left or right uniformly when passing through the
disclinations to obey the exclusion rule. Without loss of generality, we
require any particle passing a disclination to turn left, as shown in
Fig.~\ref{model}(a).  Otherwise, two velocity vectors pointing to a disclination
may merge to a single velocity vector after a time step if one of them turns
left and the other turns right, which violates the exclusion rule.

Here, we note that in the planar FHP lattice gas model, the momentum of the
system is conserved in both propagation and collision operations. However, by generalizing
the model to the spherical lattice, the direction of a velocity vector is
changed in its propagation, which requires a force on the particle. Furthermore, 
the bonds in the spherical lattice are generally not of the same length except
the special case of the icosahedron. Consequently, the total momentum of two or
three particles changes slightly after the binary or triple collision.
Therefore, unlike that on a planar lattice, an energy input is required to
realize the propagation and collision of the particles on the sphere. Such a
system may be realized in a group of active agents confined on the
sphere~\cite{marchetti2013hydrodynamics}.

We employ the Casper and Klug scheme to construct the spherical lattice as shown
in Fig.~\ref{model}(a)~\cite{caspar1962physical}. Starting from a regular
icosahedron with 12 vertices, we introduce $n-1$ points evenly on each edge to
divide each face into a number of triangles. The total number of vertices in the
resulting spherical lattice is $N=10n^2+2$. To perform the coarse-graining
procedure, we introduce another set of triangular lattice on the sphere with
the Casper and Klug scheme, and name it the coarse-grained lattice. The sphere
is therefore equipped with two sets of lattices: the original lattice where the
particles live, and the auxiliary coarse-grained lattice for calculating the
coarse-grained quantities. The coarse-grained lattice consists of $N_{cg}$ vertices and $2N_{cg}$
faces, each face containing $N/(2N_{cg})$ vertices in the original spherical
lattice. Summing over all the velocity vectors within a
triangular face in the coarse-grained lattice gives rise to the associated
coarse-grained velocity.

\section{Results and discussion}

\begin{figure}[th]
\centering
\includegraphics[width=2.2in]{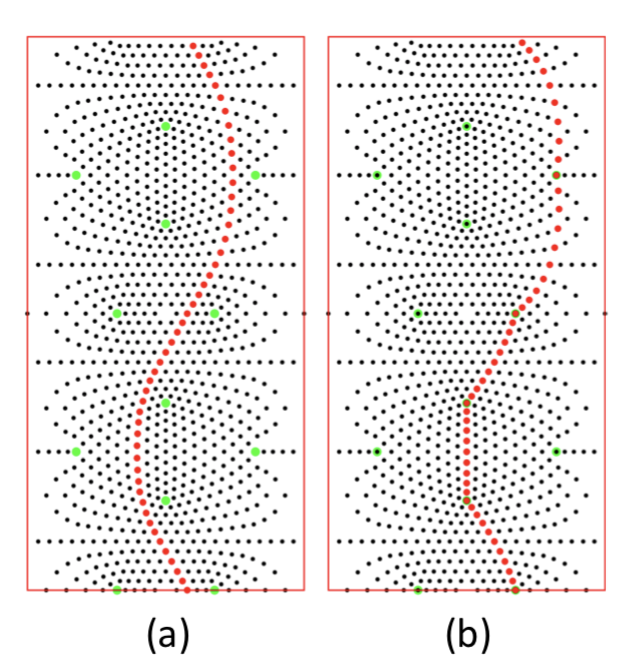}
\caption{Closed orbits (in red dots) of a single particle with different initial
  velocity directions on the spherical
  lattice plotted in the spherical coordinates. All the orbits of the particle
  with varying initial velocity directions
  are closed, and have the same period. The horizontal and vertical axes are for
$\theta$ and $\phi$, respectively.  The green dots represent the five-fold
disclinations. $N=1002$.
    } \label{periodic_motion}
\end{figure}

\begin{figure}[th]
\centering
\includegraphics[width=3.4in]{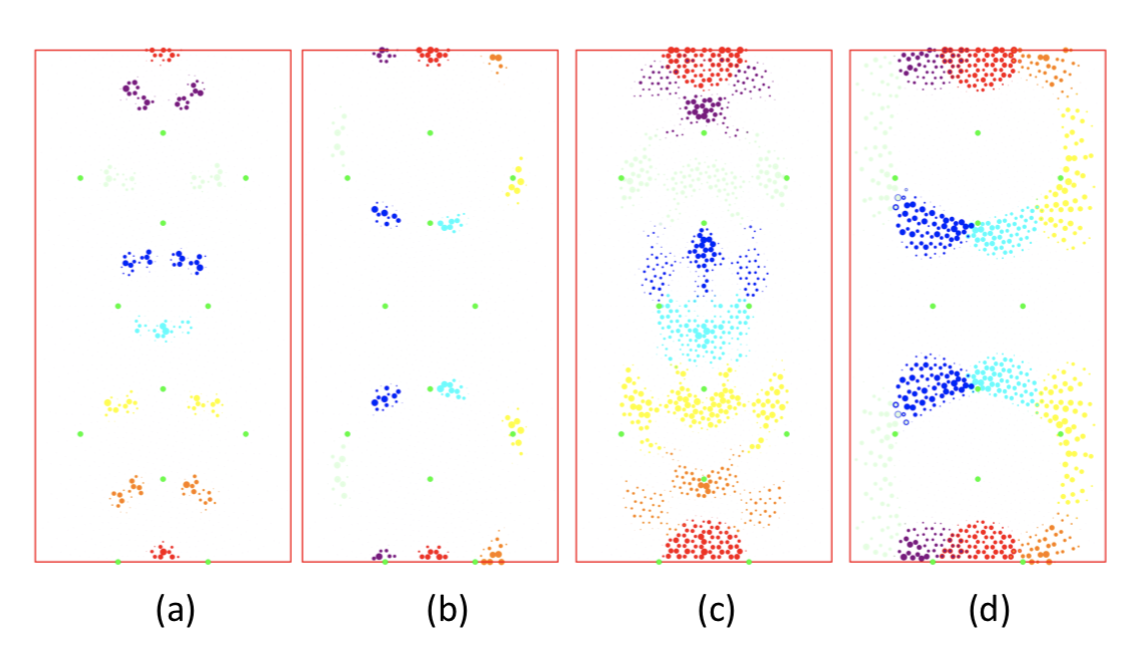}
\caption{Periodic separation and merging of the moving clusters. The
  morphologies of the cluster at different time steps are shown in different
    colors: from below to above, $t=1$ (red), $30$ (orange), $60$ (yellow), $90$
    (cyan), $120$ (light green), and $180$ (purple). The size of the colored
    dots indicates the number of particles in the coarse-grained cells. The
    green dots represent the disclinations in the original spherical lattice.
    The sizes of the clusters are $r_c=0.2 R$ (a,c) and $0.5 R$ (b,d).
    $N=16002$.  $N_{cg}=1002$. 
    } \label{moving_cluster}
\end{figure}

\begin{figure*}[th]  
\centering 
\subfigure[]{
\includegraphics[width=5.5in]{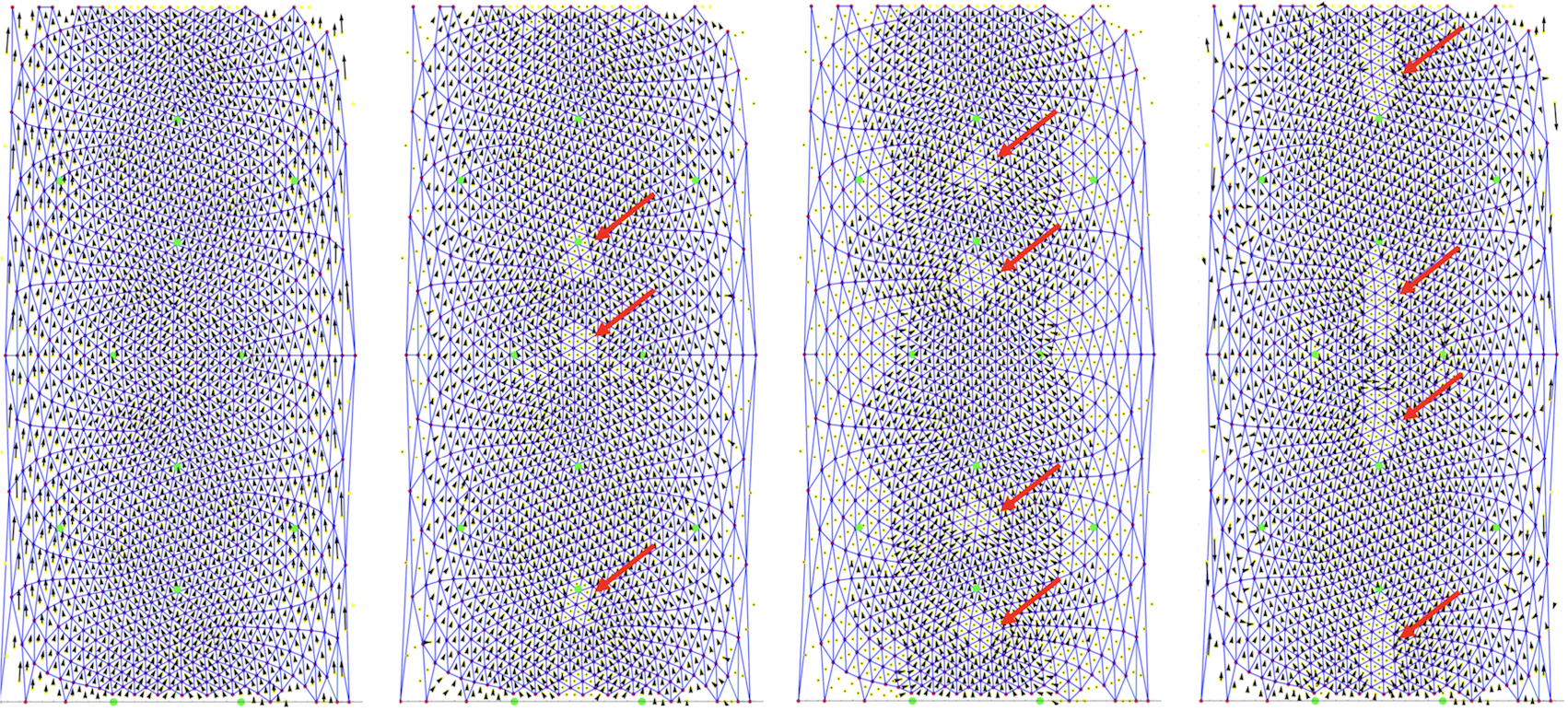}
}
\subfigure[]{
\includegraphics[width=5.5in]{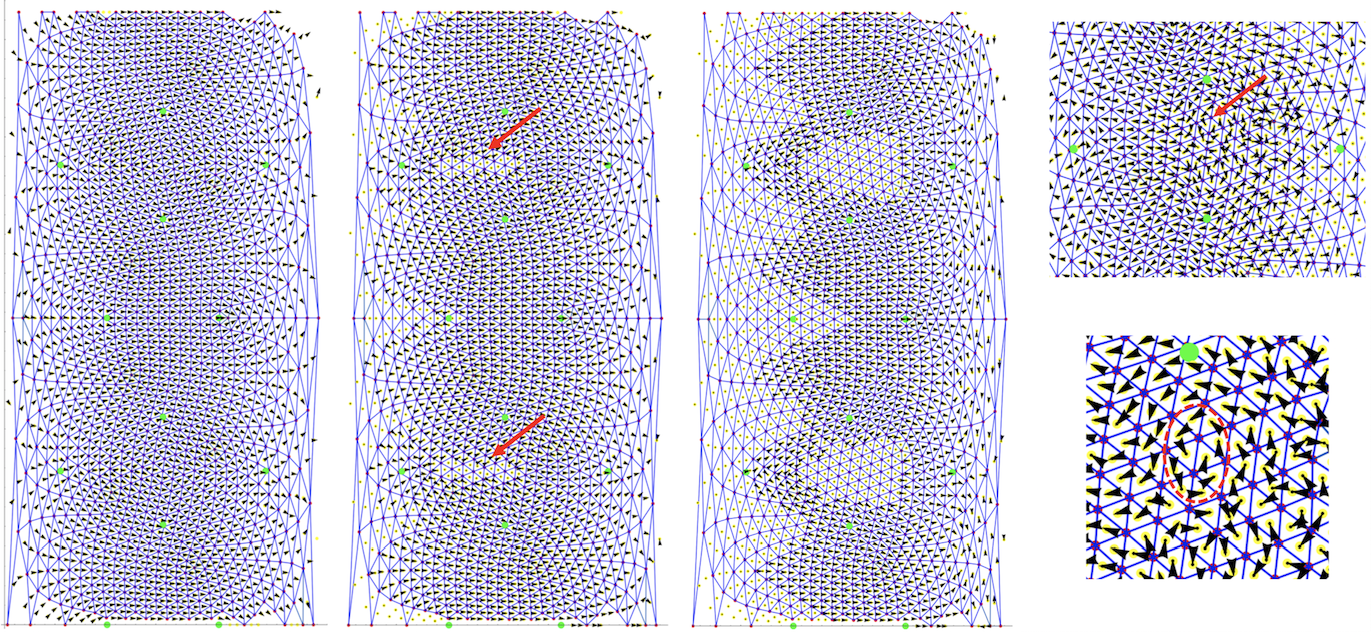}
}
\caption{The bubble and vortex structures developed in the evolution of the
  coarse-grained velocity field. Each vertex is occupied by a particle in the
    initial state, and all the particles move quasi-uniformly along
    $\hat{e}_{\phi}$ (a) and $\vec{e}_{\theta}$ (b).  (a) shows the
    transportation and deformation of the void areas (the bubbles), as indicated
    by the red arrows.  (b) shows the emergent vortex structure (indicated by
	the red arrow in the last snapshot whose zoom-in plot is shown below) in
    addition to the bubbles. The snapshots in (a) and (b) are at $t=1$, $11$,
	     $35$, $100$ (a), and $t=1$, $16$, $45$, $278$ (b), respectively.
	       The green dots represent the disclinations. $N=16002$ (a), and
	       $64002$ (b). $N_{cg}=1002$. 
} \label{bubbles}
\end{figure*}

We first consider the motion of a single particle on the spherical lattice. We map
the spherical lattice in the three-dimensional Euclidean space to the plane of the
spherical coordinates $\{ \theta, \phi \}$. $\theta \in [0, \pi]$. $\phi \in [0,
	  2\pi)$. In the rectangular box in
Fig.~\ref{periodic_motion}, the short and long sides are the axes of $\theta$
and $\phi$, respectively. The black dots are the vertices in the spherical
lattice, and the green dots represent the 12 five-fold disclinations. The trajectory
of the single particle is shown in red dots. We see that the trajectory is
always closed whether the particle passes through a disclination or not. In
particular, if the particle passes through a disclination, as shown in
Fig.~\ref{periodic_motion}(b), it also passes through other four disclinations,
forming a closed loop over the sphere. We notice that the period of any closed
orbit is the same regardless of the initial position and velocity of
the particle.

The combination of the identical period of the closed orbit and the spherical geometry
may lead to periodic morphological transformation of a collectively moving
cluster of particles. We create a moving cluster of particles on the spherical lattice to test
this conjecture.  We first select all the vertices in the spherical lattice whose
distance from any reference vertex is smaller than $r_c$, and then put
particles on these vertices.  The initial velocity of these particles is along
one of the six (or five, if a particle sits at a disclination) discrete directions
that makes the minimum angle with either the local $\hat{e}_{\phi}$ or $\hat{e}_{\theta}$
directions. Thus the cluster is equipped with a quasi-uniform velocity
distribution. The evolutions of typical clusters along $\hat{e}_{\phi}$ and
$\hat{e}_{\theta}$ are shown in Fig.~\ref{moving_cluster}. For visual
convenience, we work in the coarse-grained lattice; the size of the colored
dots indicates the number of particles in each coarse-grained cell. The
tempo-spatial patterns of the clusters are shown by different colors from red
at $t=1$ [see the lower parts in Fig.~\ref{moving_cluster}] to purple at $t=180$ [at
    the top in Fig.~\ref{moving_cluster}]. A salient feature in the cluster evolution is their periodic
separation and merging as a whole. The periodic morphological change of the
clusters is guided by the curved lattice that converges and diverges
periodically around the disclinations. This observation suggests that the curvature
may be exploited to control the morphology of the moving clusters. For example, in
a two-dimensional active matter system realizable by a group of swimming bacteria or
other active agents, one may create various geometric structures like
Gaussian bumps to control the collective motion~\cite{marchetti2013hydrodynamics}. Note
that the motions of the clusters in Fig.~\ref{moving_cluster} are collision
free for the small size of the clusters.

When the particles occupy the entire spherical lattice, all quasi-uniformly moving
along $\hat{e}_{\phi}$, their collective movement gives rise to multiple bubble
structures, as shown in Fig.~\ref{bubbles}. A frame-by-frame examination shows
that, prior to the appearance of these bubbles, local velocity vectors are found
to move in
and out along multiple directions. The irregularity of these velocity vectors
leads to the emergence of void areas embedded in the coarse-grained velocity
field.  In other words, the bubble structures emerge to avoid singularity in
the coarse-grained velocity field. Here, we emphasize that the bubbles are still filled with
particles, but the coarse-grained velocity
field vanishes therein. Note that in a larger system of $N=64002$,
where half of the bonds are occupied by particles, a few coarse-grained velocity
vectors are scattered inside some bubbles. It is observed that the bubble size
can grow up to about 20 coarse-grained cells from $t=1$ to $t=30$. In
Fig.~\ref{bubbles}(a) and \ref{bubbles}(b), we also show the collective
progression and deformation of the bubbles.

\begin{figure}[t]
\centering
\includegraphics[width=3.5in]{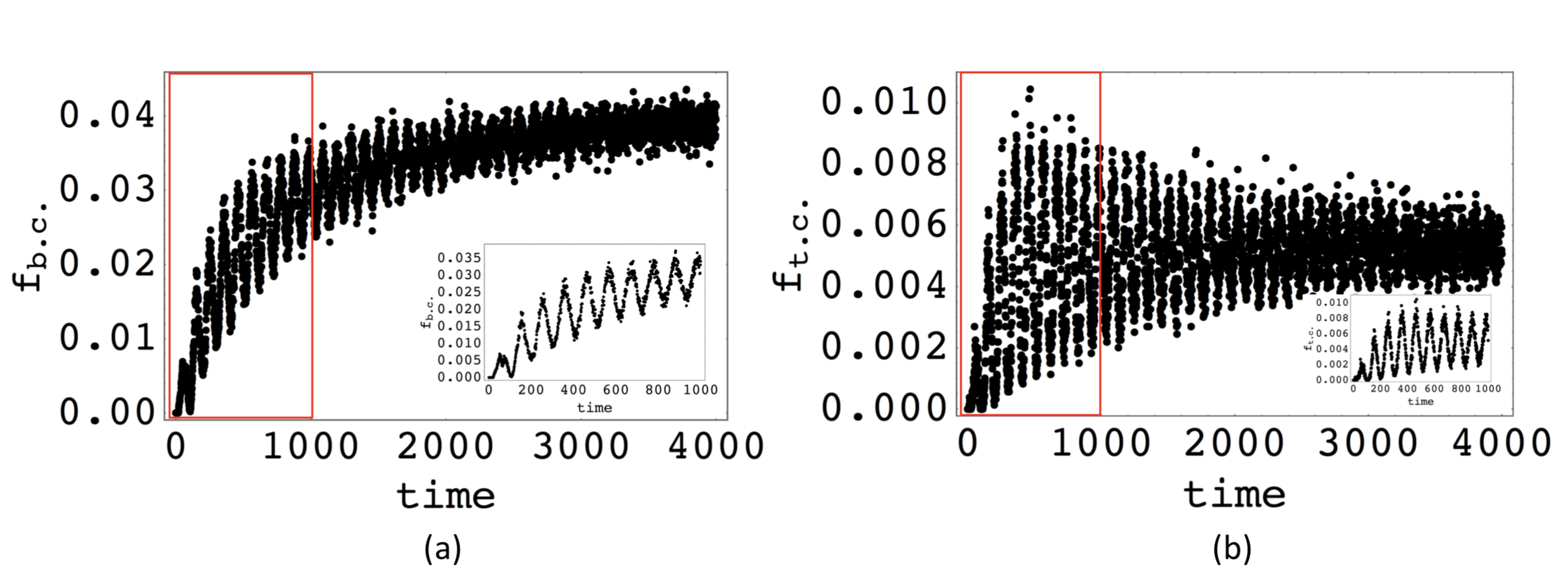}
\caption{Plot of the frequencies of the binary and triple collisions versus time.
Both curves exhibit the oscillation behaviors. The insets are the zoom-in plots of the
  curves in the red rectangles. Each vertex is occupied by a particle in the
    initial state, and all the particles move quasi-uniformly along
$\hat{e}_{\theta}$. $N=16002$.    
    } 
    \label{collision}
\end{figure}

For quasi-uniformly moving particles along $\hat{e}_{\theta}$ over the entire
spherical lattice, each occupying
one vertex, we observe vortex structures in addition to the
bubble structures, as shown in Fig.~\ref{bubbles}(b). Vortices are interesting
emergent structures in macroscopic hydrodynamic systems. An salient feature of
two-dimensional turbulence is the emergence of isolated coherent
vortices~\cite{kaneda2013statistical}. Our model allows one to observe the
formation of the vortices at the microscopic level. We track the evolution of the
coarse-grained velocity vectors and find that the vortices start to form when
the group of particles moving quasi-uniformly along $\vec{e}_{\theta}$ meet those
that have passed through the south pole and are moving along
$-\vec{e}_{\theta}$. A typical vortex structure is shown in the dashed ellipse
in the last inset in Fig.~\ref{bubbles}(b). The vortex turns out to be rather
robust.  It is largely anchored at the same place in the evolving velocity
field for a period up to $\Delta t = 10$. It is worth pointing out
that the emergence of the non-equilibrium dynamic structures of bubbles and
vortices results from the bent lattice around the disclinations instead of the
disclinations themselves.  Figure~\ref{bubbles} shows no correlation between the
locations of the disclinations and the sites of the bubbles and vortices.

With the accumulation of the collision effect, both the bubble and vortex structures
are ultimately destroyed. We track the collision events in the propagation of the
particles and plot the frequencies of the binary and triple collisions versus time
in Fig.~\ref{collision} for the case where the particles move quasi-uniformly along
$\hat{e}_{\phi}$. Here, the collision frequency refers to
the number of the collision events rescaled by the total number of particle
movements in a time step. Figure~\ref{collision} shows that both the binary and triple
collision curves exhibit the oscillation behaviors. The period of the
oscillation is the same as that of the closed orbit formed by a single
particle.  Figure~\ref{collision}(a) shows that the binary collision between
particles becomes more frequent at a large time scale, and the curve
ultimately converges towards an upper bound. In contrast, the oscillation amplitude
of the triple collision curve shrinks in time, as shown in
Fig.~\ref{collision}(b).  With the convergence of both collision curves, the
bubble and vortex structures gradually dissolve, and the system evolves
towards equilibrium.

\begin{figure}[t]  
\centering \subfigure[]{
\includegraphics[width=1.58in]{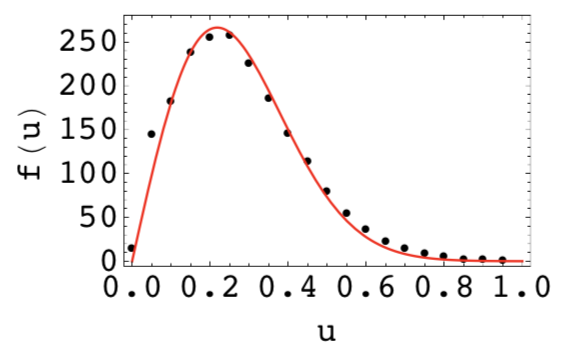}
}
\subfigure[]{
\includegraphics[width=1.58in]{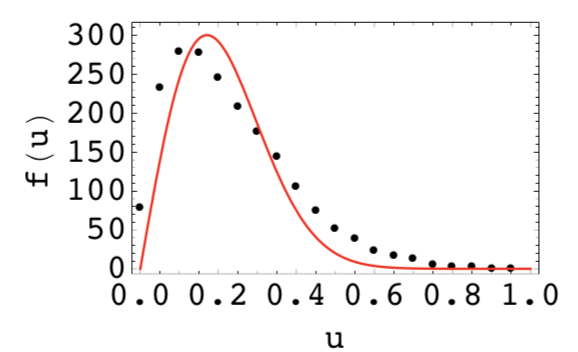}
} 
\subfigure[]{
\includegraphics[width=1.58in]{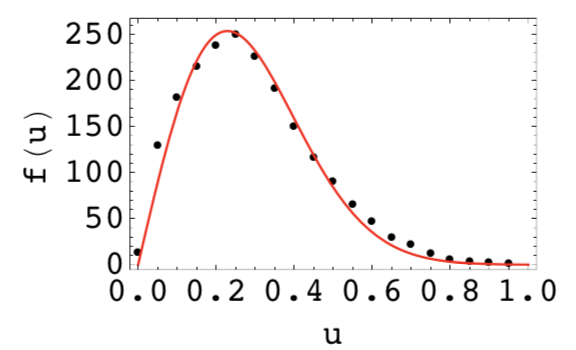}
}
\subfigure[]{
\includegraphics[width=1.58in]{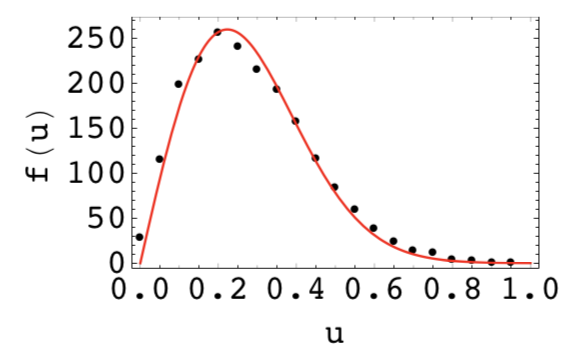}
}
\caption{Roles of the binary and triple collisions on the equilibrium
  distribution of the tempo-spatially coarse-grained speed.  The solid red
    fitting curves have the common functional form of $f_{fit}(u)=a u \exp(-b
	u^2)$. (a) with both the binary and triple collision rules. (b) the
    collision free case. (c) with only the binary collision rule. (d) with only
    the triple collision rule. The time average is over $t\in[100, 110]$. The
    speed is scaled by the maximum coarse-grained speed.  Each bond has $50\%$
    probability to be occupied by a unit velocity vector in the initial state.
    The values for the parameters in the fitting function are obtained from the
    Quasi-Newton method. (a) $a=2014.18$, $b=10.52$; (b) $a = 2898.48$, $b = 17.1363$; (c) $a =
    1827.16$, $b = 9.55697$; (d) $a = 1918.49$. $b = 10.0383$. $N=64002$.
    $N_{cg}=1002$.   
} \label{f_u12}
\end{figure}

Now, we analyze the equilibrium property of the spherical lattice gas model. 
Consider a spherical lattice where each bond in the original lattice has $50\%$
probability to be occupied.  It turns out that the influence of the initial state of the system
is significantly diminished after tens of time steps. Statistical analysis of the
tempo-spatially coarse-grained velocity field $\vec{u}(\vec{x})$ reveals a
stable distribution $f(u)$, as shown in Fig~\ref{f_u12}(a). Here, $u$ is
rescaled by its maximum value.
$\vec{u}(\vec{x})$ is defined at the center of each triangular coarse-grained
cell, and it is the sum of the velocities of the particles within the coarse-grained
cell. $f(u) \Delta u$ is the number of the coarse-grained cells whose associated
coarse-grained speed falls between $u$ and $u+\Delta u$. The data in
Fig~\ref{f_u12}(a) can be well fitted by the function $f_{fit}(u)=a u \exp(-b
u^2)$. It is recognized as the two-dimensional Maxwell-Boltzmann distribution by
expressing $b$ as $b=m/(2k_BT)$, where $k_B$ is the Boltzmann constant, $T$ is
the temperature, and $m$ is the total mass of the particles contained in each
coarse-grained cell.  Since both $m$ and $T$ are linear to the average number of
particles $\langle N_{{\rm cell}} \rangle$ in the coarse-grained cells, $b$ is
expected to be independent of the degree of the coarse graining. To check this
expectation, we simulate systems with
varying $N$. $N=\{25002, 36002, 49002, 64002, 81002\}$, and
$\langle N_{{\rm cell}} \rangle= \{36, 48, 67, 88, 112\}$, respectively. It is found that $b=10.81 \pm
0.43$ in all these systems, which is indeed almost an invariant.

To understand the emergent statistical law,
we regard each coarse-grained cell as a virtual particle. For convenience
in later discussions, the particles in the original spherical lattice
are named molecular particles. The movement of the molecular particles into and out of the
cells results in the transfer of momentum among neighboring virtual
particles. A virtual particle's momentum varies in a ``quantum" manner. The minimum
momentum change of a virtual particle is the momentum of an individual molecular
particle. The effect of the
collisions among molecular particles within a virtual particle is
to fully randomize the $N_{{\rm cell}}$ unit velocity vectors $\vec{v}_i$ ($i \in [1,
N_{{\rm cell}}]$) such that these velocity vectors are statistically independent. 
The velocity of a virtual particle $\vec{u}$ is the sum of
the associated molecular particles' velocities: 
\begin{eqnarray} 
\vec{u} = \sum_{i=1}^{N_{{\rm cell}}} \vec{v}_i.
\end{eqnarray}
It is recognized that $\vec{u}$ is essentially the sum of $N_{{\rm cell}}$
independent and identically distributed random unit vectors $\vec{v}_i$ ($i \in [1, N_{{\rm cell}}]$). Our problem is
therefore mapped to a random walk model defined on a triangular lattice.
According to the central limit theorem, the end-to-end distance $r$ of a random walker
in a triangular lattice asymptotically conforms to the Rayleigh distribution after
a large number of steps~\cite{berg1993random}: 
\begin{eqnarray} 
p_1(r; N_{{\rm cell}})= A_1 \frac{r}{N_{{\rm cell}}}
\exp \left[ -\frac{r^2}{2 r_0^2 N_{{\rm cell}}} \right],
\label{p1}
\end{eqnarray}
where $r_0$ is the lattice spacing of the triangular lattice, and $A_1$ is the
normalization coefficient. Note that Eq.(\ref{p1}) and $f_{fit}(u)$ in
Fig.~\ref{f_u12}(a) share the same functional form, and both of them correspond
to the two-dimensional Maxwell-Boltzmann distribution.

However, we find that $N_{{\rm cell}}$, the number of particles in the
coarse-grained cells, is subject to the Gaussian type
fluctuation, which will be discussed later. This population fluctuation must be
taken into account in the random walk model. In other words, the random walk
model shall be generalized to allow the fluctuation in the number of
steps. The probability density for
$N_{{\rm cell}}$ can be written as
\begin{eqnarray}
p_2(N_{{\rm cell}}) = A_2 \exp\left[ -\frac{(N_{{\rm cell}} -
\langle N_{{\rm cell}} \rangle)^2}{a_2^2} \right].
\label{p2}
\end{eqnarray}
In the following, we discuss if the population fluctuations in the coarse-grained cells
will influence the statistical law governing the speed distribution of the virtual
particles. From Eq.~(\ref{p1}) and (\ref{p2}), we have the probability density
for $u$:
\begin{eqnarray} 
p(u) = \int_0^{\infty}p_1(u; N_{{\rm cell}}) \times p_2(N_{{\rm cell}})
d N_{{\rm cell}},
\label{pu}
\end{eqnarray}
where the integration is over all possible values for $N_{{\rm cell}}$. We
notice that the dominant contribution to the integral is from the domain near
$N_{{\rm cell}}=\langle N_{{\rm cell}} \rangle$, which is at the peak of the $p_2(N_{{\rm
cell}})$ curve. Eq.(\ref{pu}) can be evaluated by the saddle
point method~\cite{arfken1999mathematical}: 
\begin{eqnarray} 
p(u) \approx \frac{A a_2 u}{\langle N_{{\rm cell}}
\rangle} \exp\left[  -\frac{u^2}{2 u_0^2 \langle N_{{\rm cell}} \rangle} \right], 
\label{pu_saddle}
\end{eqnarray}
where $A$ is the normalization coefficient, and $u_0$ is the magnitude of the
molecular velocity vector. By comparing Eq.(\ref{pu_saddle}) with the numerical evaluation of
Eq.(\ref{pu}), we find that Eq.(\ref{pu_saddle}) is a good approximation even
when the standard deviation in $N_{{\rm cell}}$ is as large as
about $30\%$ around its average value. To conclude, modest population fluctuation
in a virtual particle does not change the
nature of the statistical law for the distribution of the coarse-grained speed $u$.

In mapping our problem to the generalized random walk model, an
important assumption is the collision driven randomization of the molecular
velocity vectors. Now we analyze the specific roles of the binary and triple
collisions in achieving the regular statistical distribution. We first remove
both types of collisions from our model and examine the consequence. Figure~\ref{f_u12}(b)
shows the distribution of the coarse-grained speed $u$ without any collision. We see that the simulation
data are deviated from the two-dimensional Maxwell-Boltzmann distribution. The
peak on the curve in Fig.~\ref{f_u12}(b) is due to the random initial state of the
molecular particles. The result presented in Fig.~\ref{f_u12}(b) indicates that the
randomness brought by the initial condition is insufficient to establish a
Maxwell-Boltzmann distribution.

To clarify if the solo binary collision suffices to establish the statistical
regularity in the speed distribution, we remove the triple collision from the
model and keep only the binary collision rule. In Fig~\ref{f_u12}(c), we show
that the binary collision seems sufficient to destroy the statistical correlation
of molecular particles' velocities, and produce the result of the random walk
model. We further find that only retaining the triple collision rule in our model
also suffices to randomize the velocities, as suggested by the curve in Fig.~\ref{f_u12}(d).
Note that the triple collision typically occurs at much lower rate than the binary
collision. We observe $50\%$ more binary collision events than the triple collision
events in the system of about 30000 molecular particles. These observations
suggest that even a relatively rare collision rate between particles can
randomize the molecular particles and produce a regular statistical
distribution. We also study systems of higher particle densities. It turns out
that the collision rate is even lower. In the spherical
lattice where $80\%$ of the bonds are occupied and $N=100002$, $f_{b.c.} =
0.06\%$ and $f_{t.c.} = 0.15\%$, which are only $4\%$ and $15\%$ of
the corresponding collision frequencies for the case of $50\%$ occupation rate. 
Even at such a low collision rate, we still observe that the statistics of the
coarse-grained speed obeys the Maxwell-Boltzmann distribution.

\begin{figure}[t]  
\centering 
\subfigure[]{
\includegraphics[width=1.58in]{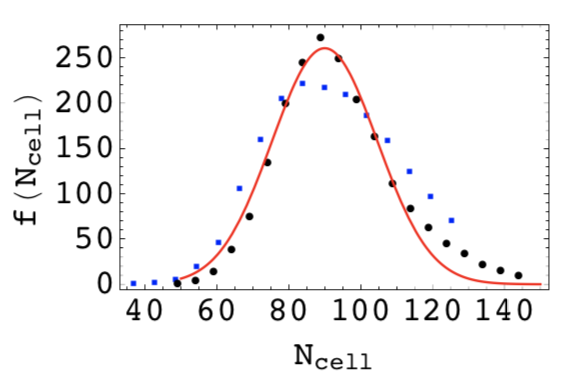}
}
\subfigure[]{
\includegraphics[width=1.58in]{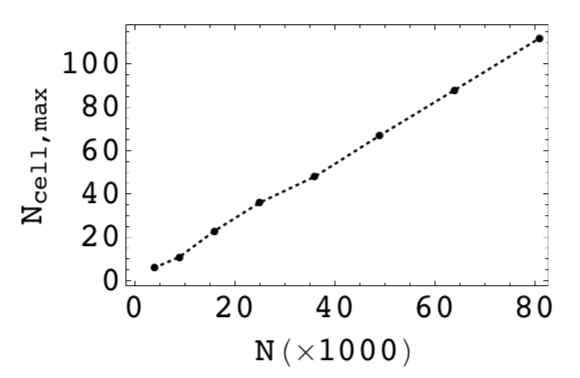}
}
\caption{Analysis of the population fluctuation in the triangular cells of the
  coarse-grained spherical lattice.  (a) shows the distribution of the number of
    particles $N_{{\rm cell}}$ in each cell. The square blue dots are for the
    collision free case, and the black dots are for the case with both the
    binary and triple collision rules. The black dots, especially those near the maximum $N_{{\rm
      cell}}$, are fitted by the Gaussian distribution function. The
      time average is over $\Delta t=10$.   $N=64002$.  $N_{cg}=1002$.
(b) Shows the linear dependence of the maximum $N_{{\rm cell}}$ on the number of
vertices $N$ in the spherical lattice. The total number of molecular particles
is half of $N$. Both the
    binary and triple collision rules are applied. 
} \label{Ncell_plot}
\end{figure}

Collisions between molecular particles are also found to be responsible for the
regular statistical distribution of $N_{{\rm cell}}$, the number of molecular
particles in the coarse-grained cells.
Statistical analysis of $N_{{\rm cell}}$ in each coarse-grained cell reveals
that the distribution of $N_{{\rm cell}}$ approaches the Gaussian distribution by adding the
collision rules, as illustrated in Fig.~\ref{Ncell_plot} (a). In
Fig.~\ref{Ncell_plot}(b), we show the linear relation between $N_{{\rm cell,
max}}$ [the value of $N_{{\rm cell}}$ at the peak of the $f(N_{{\rm cell}})$
curve] and the total number of molecular particles which is half of $N$.
Simulations show that the distribution of the speed $u$ of the virtual particles
is well fitted by the
Maxwell-Boltzmann distribution when $\langle N_{{\rm cell}} \rangle$ is
larger than about 40.

Finally, we exploit the fact that the emergent statistical law of the virtual
particles' speed is independent of the number of molecular particles $N_{{\rm
  cell}}$ within each virtual particle once $N_{{\rm cell}}$ is sufficiently
  large.  The coarse-graining procedure maps the molecular particles into a
  collection of virtual particles. The velocity of an arbitrary virtual particle labelled
  $j$ is  
\begin{eqnarray}
\vec{u}^{(1)}_j=\sum_{i\in \Omega_j^{(1)}}\vec{v}_i.
\label{u1}
\end{eqnarray}
$\vec{v_i}$ is the velocity of the molecular particle labelled $i$ within the
virtual particle $j$. It is a random unit vector whose orientation can take one
of the six discrete directions in the triangular lattice with the same
probability. The sum in Eq.(\ref{u1}) is over all the molecular particles
in the coarse-grained cell $\Omega_j^{(1)}$ that is named as the virtual particle
$j$. The
superscript $``(1)"$ in $\vec{u}^{(1)}_j$ and $\Omega_j^{(1)}$ denotes the
first round coarse-graining
procedure from the original molecular particles to the virtual particles. We have shown that the
magnitude of the
resulting random variable $\vec{u}^{(1)}$ conforms to the two-dimensional Maxwell-Boltzmann
distribution. One can perform further coarse graining by classifying all the virtual
particles in the newly built coarse-grained cells denoted as $\Omega_k^{(2)}$. Summing over
all the virtual particles' velocities in $\Omega_k^{(2)}$ returns a new random
variable:
\begin{eqnarray}
\vec{u}^{(2)}_k=\sum_{i\in \Omega_k^{(2)}} \vec{u}_i^{(1)}.
\label{u2}
\end{eqnarray}
The comparison of Eq.(\ref{u1}) and (\ref{u2}) shows that the
magnitude of $\vec{u}^{(2)}$ also obeys the two-dimensional
Maxwell-Boltzmann distribution. The coarse-graining procedure can be further
extended to higher levels. On the premise that there are sufficient virtual particles in the
coarse-grained cells, the statistical law always holds. The transferable statistical law from the
original molecular particle system to any coarse-grained system is essentially
attributed to the statistical independence of the molecular particles at the
very bottom level; no statistical correlation can be built up in the
coarse-graining process itself.

\section{CONCLUSION}

In summary, based on the spherical lattice gas automaton with the well-defined
propagation and collision rules, we reveal several emergent dynamic structures
resulting from the geometric frustration of the collectively moving clusters,
including the periodic global morphological transformations, the bubble and vortex
structures embedded in the coarse-grained velocity field. With the accumulation
of the collision effect, statistical analysis shows that the coarse-grained speed
conforms to the two-dimensional Maxwell-Boltzmann distribution despite the
modest population fluctuations. A generalized random walk model is proposed to
account for the observed regular statistical behaviors. This work demonstrates
the capability of the spherical lattice gas automaton in revealing the non-equilibrium dynamic
structures and in producing the statistical
law of the real gas. For its simplicity and the capability of simulating physics, the
lattice gas automaton model may be generalized to other curved surfaces to
explore geometrically driven non-equilibrium physics. Furthermore, considering
that the emergent dynamic structures like the bubbles and vortices in our system
are largely driven by the disclinations in the lattice, it is of interest to
explore the potential collective dynamics regulated by a designed mix of five-
and seven-fold disclinations on an almost-planar lattice.

\section*{Acknowledgement}

This work was supported by NSFC Grants No. 16Z103010253, the SJTU startup fund
under Grant No. WF220441904, and the award of the Chinese Thousand Talents
Program for Distinguished Young Scholars under Grant No. 16Z127060004.


\begin{thebibliography}{37}
\expandafter\ifx\csname natexlab\endcsname\relax\def\natexlab#1{#1}\fi
\expandafter\ifx\csname bibnamefont\endcsname\relax
  \def\bibnamefont#1{#1}\fi
\expandafter\ifx\csname bibfnamefont\endcsname\relax
  \def\bibfnamefont#1{#1}\fi
\expandafter\ifx\csname citenamefont\endcsname\relax
  \def\citenamefont#1{#1}\fi
\expandafter\ifx\csname url\endcsname\relax
  \def\url#1{\texttt{#1}}\fi
\expandafter\ifx\csname urlprefix\endcsname\relax\def\urlprefix{URL }\fi
\providecommand{\bibinfo}[2]{#2}
\providecommand{\eprint}[2][]{\url{#2}}

\bibitem[{\citenamefont{Lee and Yang}(1952)}]{lee1952statistical}
\bibinfo{author}{\bibfnamefont{T.-D.} \bibnamefont{Lee}} \bibnamefont{and}
  \bibinfo{author}{\bibfnamefont{C.-N.} \bibnamefont{Yang}},
  \bibinfo{journal}{Phys. Rev.} \textbf{\bibinfo{volume}{87}},
  \bibinfo{pages}{410} (\bibinfo{year}{1952}).

\bibitem[{\citenamefont{Baxter}(1982)}]{baxter1982exactly}
\bibinfo{author}{\bibfnamefont{R.~J.} \bibnamefont{Baxter}},
  \emph{\bibinfo{title}{Exactly solved models in statistical mechanics}}
  (\bibinfo{publisher}{Elsevier}, \bibinfo{year}{1982}).

\bibitem[{\citenamefont{Kadanoff}(1986)}]{kadanoff1986two}
\bibinfo{author}{\bibfnamefont{L.~P.} \bibnamefont{Kadanoff}},
  \bibinfo{journal}{Physics Today} \textbf{\bibinfo{volume}{39}},
  \bibinfo{pages}{7} (\bibinfo{year}{1986}).

\bibitem[{\citenamefont{Kadanoff}(1999)}]{kadanoff1999order}
\bibinfo{author}{\bibfnamefont{L.~P.} \bibnamefont{Kadanoff}},
  \emph{\bibinfo{title}{From order to chaos II: Essays: Critical, chaotic and
  otherwise}} (\bibinfo{publisher}{World Scientific}, \bibinfo{year}{1999}).

\bibitem[{\citenamefont{Bagnoli et~al.}(2001)\citenamefont{Bagnoli, Boccara,
  and Rechtman}}]{bagnoli2001nature}
\bibinfo{author}{\bibfnamefont{F.}~\bibnamefont{Bagnoli}},
  \bibinfo{author}{\bibfnamefont{N.}~\bibnamefont{Boccara}}, \bibnamefont{and}
  \bibinfo{author}{\bibfnamefont{R.}~\bibnamefont{Rechtman}},
  \bibinfo{journal}{Phys. Rev. E} \textbf{\bibinfo{volume}{63}},
  \bibinfo{pages}{046116} (\bibinfo{year}{2001}).

\bibitem[{\citenamefont{Bagnoli et~al.}(2005)\citenamefont{Bagnoli, Franci, and
  Rechtman}}]{bagnoli2005phase}
\bibinfo{author}{\bibfnamefont{F.}~\bibnamefont{Bagnoli}},
  \bibinfo{author}{\bibfnamefont{F.}~\bibnamefont{Franci}}, \bibnamefont{and}
  \bibinfo{author}{\bibfnamefont{R.}~\bibnamefont{Rechtman}},
  \bibinfo{journal}{Physical Review E} \textbf{\bibinfo{volume}{71}},
  \bibinfo{pages}{046108} (\bibinfo{year}{2005}).

\bibitem[{\citenamefont{Preston~Jr and Duff}(2013)}]{preston2013modern}
\bibinfo{author}{\bibfnamefont{K.}~\bibnamefont{Preston~Jr}} \bibnamefont{and}
  \bibinfo{author}{\bibfnamefont{M.~J.} \bibnamefont{Duff}},
  \emph{\bibinfo{title}{Modern cellular automata: theory and applications}}
  (\bibinfo{publisher}{Springer Science \& Business Media},
  \bibinfo{year}{2013}).

\bibitem[{\citenamefont{Hardy et~al.}(1973)\citenamefont{Hardy, Pomeau, and
  De~Pazzis}}]{hardy1973time}
\bibinfo{author}{\bibfnamefont{J.}~\bibnamefont{Hardy}},
  \bibinfo{author}{\bibfnamefont{Y.}~\bibnamefont{Pomeau}}, \bibnamefont{and}
  \bibinfo{author}{\bibfnamefont{O.}~\bibnamefont{De~Pazzis}},
  \bibinfo{journal}{J. Math. Phys.} \textbf{\bibinfo{volume}{14}},
  \bibinfo{pages}{1746} (\bibinfo{year}{1973}).

\bibitem[{\citenamefont{Hardy et~al.}(1976)\citenamefont{Hardy, De~Pazzis, and
  Pomeau}}]{hardy1976molecular}
\bibinfo{author}{\bibfnamefont{J.}~\bibnamefont{Hardy}},
  \bibinfo{author}{\bibfnamefont{O.}~\bibnamefont{De~Pazzis}},
  \bibnamefont{and} \bibinfo{author}{\bibfnamefont{Y.}~\bibnamefont{Pomeau}},
  \bibinfo{journal}{Phys. Rev. A} \textbf{\bibinfo{volume}{13}},
  \bibinfo{pages}{1949} (\bibinfo{year}{1976}).

\bibitem[{\citenamefont{Margolus et~al.}(1986)\citenamefont{Margolus, Toffoli,
  and Vichniac}}]{margolus1986cellular}
\bibinfo{author}{\bibfnamefont{N.}~\bibnamefont{Margolus}},
  \bibinfo{author}{\bibfnamefont{T.}~\bibnamefont{Toffoli}}, \bibnamefont{and}
  \bibinfo{author}{\bibfnamefont{G.}~\bibnamefont{Vichniac}},
  \bibinfo{journal}{Phys. Rev. Lett.} \textbf{\bibinfo{volume}{56}},
  \bibinfo{pages}{1694} (\bibinfo{year}{1986}).

\bibitem[{\citenamefont{Frisch and Pomeau}(1986)}]{Frisch1986}
\bibinfo{author}{\bibfnamefont{U.}~\bibnamefont{Frisch}},
  \bibinfo{author}{\bibfnamefont{B.}~\bibnamefont{Hasslacher}},
  \bibnamefont{and}
  \bibinfo{author}{\bibfnamefont{Y.}~\bibnamefont{Pomeau}},
  \bibinfo{journal}{Phys. Rev. Lett.} \textbf{\bibinfo{volume}{56}},
  \bibinfo{pages}{1505} (\bibinfo{year}{1986}).

\bibitem[{\citenamefont{Frisch et~al.}(1987)\citenamefont{Frisch,
  d$^\prime$Humieres, Hasslacher, Lallemand, Pomeau, Rivet
  et~al.}}]{frisch1987lattice}
\bibinfo{author}{\bibfnamefont{U.}~\bibnamefont{Frisch}},
  \bibinfo{author}{\bibfnamefont{D.}~\bibnamefont{d$^\prime$Humieres}},
  \bibinfo{author}{\bibfnamefont{B.}~\bibnamefont{Hasslacher}},
  \bibinfo{author}{\bibfnamefont{P.}~\bibnamefont{Lallemand}},
  \bibinfo{author}{\bibfnamefont{Y.}~\bibnamefont{Pomeau}},
  \bibinfo{author}{\bibfnamefont{J.-P.} \bibnamefont{Rivet}},
  \bibnamefont{et~al.}, \bibinfo{journal}{Complex Systems}
  \textbf{\bibinfo{volume}{1}}, \bibinfo{pages}{649} (\bibinfo{year}{1987}).

\bibitem[{\citenamefont{Swift et~al.}(1996)\citenamefont{Swift, Orlandini,
  Osborn, and Yeomans}}]{swift1996lattice}
\bibinfo{author}{\bibfnamefont{M.~R.} \bibnamefont{Swift}},
  \bibinfo{author}{\bibfnamefont{E.}~\bibnamefont{Orlandini}},
  \bibinfo{author}{\bibfnamefont{W.R.}~\bibnamefont{Osborn}}, \bibnamefont{and}
  \bibinfo{author}{\bibfnamefont{J.M.}~\bibnamefont{Yeomans}},
  \bibinfo{journal}{Phys. Rev. E} \textbf{\bibinfo{volume}{54}},
  \bibinfo{pages}{5041} (\bibinfo{year}{1996}).

\bibitem[{\citenamefont{Buick and Greated}(2000)}]{buick2000gravity}
\bibinfo{author}{\bibfnamefont{J.M.}~\bibnamefont{Buick}} \bibnamefont{and}
  \bibinfo{author}{\bibfnamefont{C.A.}~\bibnamefont{Greated}},
  \bibinfo{journal}{Phys. Rev. E} \textbf{\bibinfo{volume}{61}},
  \bibinfo{pages}{5307} (\bibinfo{year}{2000}).

\bibitem[{\citenamefont{Wolf-Gladrow}(2004)}]{wolf2004lattice}
\bibinfo{author}{\bibfnamefont{D.~A.} \bibnamefont{Wolf-Gladrow}},
  \emph{\bibinfo{title}{Lattice-gas cellular automata and lattice Boltzmann
  models: an introduction}} (\bibinfo{publisher}{Springer},
  \bibinfo{year}{2004}).

\bibitem[{\citenamefont{Wolfram et~al.}(1984)}]{wolfram1984cellular}
\bibinfo{author}{\bibfnamefont{S.}~\bibnamefont{Wolfram}} \bibnamefont{et~al.},
  \bibinfo{journal}{Nature} \textbf{\bibinfo{volume}{311}},
  \bibinfo{pages}{419} (\bibinfo{year}{1984}).

\bibitem[{\citenamefont{Wolfram et~al.}(1986)}]{wolfram1986theory}
\bibinfo{author}{\bibfnamefont{S.}~\bibnamefont{Wolfram}} \bibnamefont{et~al.},
  \emph{\bibinfo{title}{Theory and applications of cellular automata}},
  vol.~\bibinfo{volume}{1} (\bibinfo{publisher}{World scientific Singapore},
  \bibinfo{year}{1986}).

\bibitem[{\citenamefont{Chen et~al.}(1995)\citenamefont{Chen, Dawson, Doolen,
  Janecky, and Lawniczak}}]{chen1995lattice}
\bibinfo{author}{\bibfnamefont{S.}~\bibnamefont{Chen}},
  \bibinfo{author}{\bibfnamefont{S.}~\bibnamefont{Dawson}},
  \bibinfo{author}{\bibfnamefont{G.}~\bibnamefont{Doolen}},
  \bibinfo{author}{\bibfnamefont{D.}~\bibnamefont{Janecky}}, \bibnamefont{and}
  \bibinfo{author}{\bibfnamefont{A.}~\bibnamefont{Lawniczak}},
  \bibinfo{journal}{Comput. Chem. Eng.} \textbf{\bibinfo{volume}{19}},
  \bibinfo{pages}{617} (\bibinfo{year}{1995}).

\bibitem[{\citenamefont{Wootton}(2001)}]{wootton2001local}
\bibinfo{author}{\bibfnamefont{J.~T.} \bibnamefont{Wootton}},
  \bibinfo{journal}{Nature} \textbf{\bibinfo{volume}{413}},
  \bibinfo{pages}{841} (\bibinfo{year}{2001}).

\bibitem[{\citenamefont{Deutsch and Dormann}(2007)}]{deutsch2007cellular}
\bibinfo{author}{\bibfnamefont{A.}~\bibnamefont{Deutsch}} \bibnamefont{and}
  \bibinfo{author}{\bibfnamefont{S.}~\bibnamefont{Dormann}},
  \emph{\bibinfo{title}{Cellular automaton modeling of biological pattern
  formation: characterization, applications, and analysis}}
  (\bibinfo{publisher}{Springer Science \& Business Media},
  \bibinfo{year}{2007}).

\bibitem[{\citenamefont{Kapral and Showalter}(2012)}]{kapral2012chemical}
\bibinfo{author}{\bibfnamefont{R.}~\bibnamefont{Kapral}} \bibnamefont{and}
  \bibinfo{author}{\bibfnamefont{K.}~\bibnamefont{Showalter}},
  \emph{\bibinfo{title}{Chemical waves and patterns}},
  vol.~\bibinfo{volume}{10} (\bibinfo{publisher}{Springer Science \& Business
  Media}, \bibinfo{year}{2012}).

\bibitem[{\citenamefont{Koelman}(1990)}]{koelman1990cellular}
\bibinfo{author}{\bibfnamefont{J.~V.~A.} \bibnamefont{Koelman}},
  \bibinfo{journal}{Phys. Rev. Lett.} \textbf{\bibinfo{volume}{64}},
  \bibinfo{pages}{1915} (\bibinfo{year}{1990}).

\bibitem[{\citenamefont{Manneville et~al.}(2012)\citenamefont{Manneville,
  Boccara, Vichniac, and Bidaux}}]{manneville2012cellular}
\bibinfo{author}{\bibfnamefont{P.}~\bibnamefont{Manneville}},
  \bibinfo{author}{\bibfnamefont{N.}~\bibnamefont{Boccara}},
  \bibinfo{author}{\bibfnamefont{G.~Y.} \bibnamefont{Vichniac}},
  \bibnamefont{and} \bibinfo{author}{\bibfnamefont{R.}~\bibnamefont{Bidaux}},
  \emph{\bibinfo{title}{Cellular Automata and Modeling of Complex Physical
  Systems: Proceedings of the Winter School, Les Houches, France, February
  21--28, 1989}}, vol.~\bibinfo{volume}{46} (\bibinfo{publisher}{Springer
  Science \& Business Media}, \bibinfo{year}{2012}).

\bibitem[{\citenamefont{Kier et~al.}(2002)\citenamefont{Kier, Cheng, and
  Testa}}]{kier2002cellular}
\bibinfo{author}{\bibfnamefont{L.~B.} \bibnamefont{Kier}},
  \bibinfo{author}{\bibfnamefont{C.-K.} \bibnamefont{Cheng}}, \bibnamefont{and}
  \bibinfo{author}{\bibfnamefont{B.}~\bibnamefont{Testa}}, \bibinfo{journal}{J.
  Theor. Biol.} \textbf{\bibinfo{volume}{215}}, \bibinfo{pages}{415}
  (\bibinfo{year}{2002}).

\bibitem[{\citenamefont{Kurka}(2012)}]{kuurka2012topological}
\bibinfo{author}{\bibfnamefont{P.}~\bibnamefont{Kurka}}, in
  \emph{\bibinfo{booktitle}{Computational Complexity}}
  (\bibinfo{publisher}{Springer}, \bibinfo{year}{2012}), pp.
  \bibinfo{pages}{3212--3233}.

\bibitem[{\citenamefont{Von~Neumann et~al.}(1966)\citenamefont{Von~Neumann,
  Burks et~al.}}]{von1966theory}
\bibinfo{author}{\bibfnamefont{J.}~\bibnamefont{Von~Neumann}},
  \bibinfo{author}{\bibfnamefont{A.~W.} \bibnamefont{Burks}},
  \bibnamefont{et~al.}, \bibinfo{journal}{IEEE Transactions on Neural Networks}
  \textbf{\bibinfo{volume}{5}}, \bibinfo{pages}{3} (\bibinfo{year}{1966}).

\bibitem[{\citenamefont{Wolfram}(1983)}]{wolfram1983statistical}
\bibinfo{author}{\bibfnamefont{S.}~\bibnamefont{Wolfram}},
  \bibinfo{journal}{Rev. Mod. Phys.} \textbf{\bibinfo{volume}{55}},
  \bibinfo{pages}{601} (\bibinfo{year}{1983}).

\bibitem[{\citenamefont{Vichniac}(1984)}]{vichniac1984simulating}
\bibinfo{author}{\bibfnamefont{G.~Y.} \bibnamefont{Vichniac}},
  \bibinfo{journal}{Physica D} \textbf{\bibinfo{volume}{10}},
  \bibinfo{pages}{96} (\bibinfo{year}{1984}).

\bibitem[{\citenamefont{Toffoli}(1984)}]{toffoli1984cellular}
\bibinfo{author}{\bibfnamefont{T.}~\bibnamefont{Toffoli}},
  \bibinfo{journal}{Physica D} \textbf{\bibinfo{volume}{10}},
  \bibinfo{pages}{117} (\bibinfo{year}{1984}).

\bibitem[{\citenamefont{Cowburn and Welland}(2000)}]{cowburn2000room}
\bibinfo{author}{\bibfnamefont{R.}~\bibnamefont{Cowburn}} \bibnamefont{and}
  \bibinfo{author}{\bibfnamefont{M.}~\bibnamefont{Welland}},
  \bibinfo{journal}{Science} \textbf{\bibinfo{volume}{287}},
  \bibinfo{pages}{1466} (\bibinfo{year}{2000}).

\bibitem[{\citenamefont{Feynman}(1982)}]{feynman1982simulating}
\bibinfo{author}{\bibfnamefont{R.~P.} \bibnamefont{Feynman}},
  \bibinfo{journal}{Int. J. Theor. Phys.} \textbf{\bibinfo{volume}{21}},
  \bibinfo{pages}{467} (\bibinfo{year}{1982}).

\bibitem[{\citenamefont{Keber et~al.}(2014)\citenamefont{Keber, Loiseau,
  Sanchez, DeCamp, Giomi, Bowick, Marchetti, Dogic, and
  Bausch}}]{keber2014topology}
\bibinfo{author}{\bibfnamefont{F.~C.} \bibnamefont{Keber}},
  \bibinfo{author}{\bibfnamefont{E.}~\bibnamefont{Loiseau}},
  \bibinfo{author}{\bibfnamefont{T.}~\bibnamefont{Sanchez}},
  \bibinfo{author}{\bibfnamefont{S.~J.} \bibnamefont{DeCamp}},
  \bibinfo{author}{\bibfnamefont{L.}~\bibnamefont{Giomi}},
  \bibinfo{author}{\bibfnamefont{M.~J.} \bibnamefont{Bowick}},
  \bibinfo{author}{\bibfnamefont{M.~C.} \bibnamefont{Marchetti}},
  \bibinfo{author}{\bibfnamefont{Z.}~\bibnamefont{Dogic}}, \bibnamefont{and}
  \bibinfo{author}{\bibfnamefont{A.~R.} \bibnamefont{Bausch}},
  \bibinfo{journal}{Science} \textbf{\bibinfo{volume}{345}},
  \bibinfo{pages}{1135} (\bibinfo{year}{2014}).

\bibitem[{\citenamefont{Marchetti et~al.}(2013)\citenamefont{Marchetti, Joanny,
  Ramaswamy, Liverpool, Prost, Rao, and Simha}}]{marchetti2013hydrodynamics}
\bibinfo{author}{\bibfnamefont{M.}~\bibnamefont{Marchetti}},
  \bibinfo{author}{\bibfnamefont{J.}~\bibnamefont{Joanny}},
  \bibinfo{author}{\bibfnamefont{S.}~\bibnamefont{Ramaswamy}},
  \bibinfo{author}{\bibfnamefont{T.}~\bibnamefont{Liverpool}},
  \bibinfo{author}{\bibfnamefont{J.}~\bibnamefont{Prost}},
  \bibinfo{author}{\bibfnamefont{M.}~\bibnamefont{Rao}}, \bibnamefont{and}
  \bibinfo{author}{\bibfnamefont{R.~A.} \bibnamefont{Simha}},
  \bibinfo{journal}{Rev. Mod. Phys.} \textbf{\bibinfo{volume}{85}},
  \bibinfo{pages}{1143} (\bibinfo{year}{2013}).

\bibitem[{\citenamefont{Yao}(2016)}]{yao2016dressed}
\bibinfo{author}{\bibfnamefont{Z.}~\bibnamefont{Yao}}, \bibinfo{journal}{Soft
  Matter} \textbf{\bibinfo{volume}{12}}, \bibinfo{pages}{7020}
  (\bibinfo{year}{2016}).


  \bibitem[{\citenamefont{Chaikin and Lubensky}(2000)}]{chaikin2000principles}
\bibinfo{author}{\bibfnamefont{P.~M.} \bibnamefont{Chaikin}} \bibnamefont{and}
  \bibinfo{author}{\bibfnamefont{T.~C.} \bibnamefont{Lubensky}},
  \emph{\bibinfo{title}{Principles of Condensed Matter Physics}},
  vol.~\bibinfo{volume}{1} (\bibinfo{publisher}{Cambridge Univ Press},
  \bibinfo{year}{2000}).



\bibitem[{\citenamefont{Casper}(1962)}]{caspar1962physical}
\bibinfo{author}{\bibfnamefont{D.}~\bibnamefont{Casper}} \bibnamefont{and}
\bibinfo{author}{\bibfnamefont{A.} \bibnamefont{Klug}}, \bibinfo{journal}{Cold
  Spring Harbor Symposia on Quantitative Biology} \textbf{\bibinfo{volume}{27}},
  \bibinfo{pages}{1-24}
  (\bibinfo{year}{1962}).


  
\bibitem[{\citenamefont{Struik}(1988)}]{struik88a}
\bibinfo{author}{\bibfnamefont{D.}~\bibnamefont{Struik}},
  \emph{\bibinfo{title}{Lectures on Classical Differential Geometry}}
  (\bibinfo{publisher}{Dover Publications}, \bibinfo{year}{1988}),
  \bibinfo{edition}{2nd} ed.




\bibitem[{\citenamefont{Kaneda and Gotoh}(2013)}]{kaneda2013statistical}
\bibinfo{author}{\bibfnamefont{Y.}~\bibnamefont{Kaneda}} \bibnamefont{and}
  \bibinfo{author}{\bibfnamefont{T.}~\bibnamefont{Gotoh}},
  \emph{\bibinfo{title}{Statistical Theories and Computational Approaches to
  Turbulence: Modern Perspectives and Applications to Global-scale Flows}}
  (\bibinfo{publisher}{Springer Science \& Business Media},
  \bibinfo{year}{2013}).

\bibitem[{\citenamefont{Berg}(1993)}]{berg1993random}
\bibinfo{author}{\bibfnamefont{H.~C.} \bibnamefont{Berg}},
  \emph{\bibinfo{title}{Random walks in biology}}
  (\bibinfo{publisher}{Princeton University Press}, \bibinfo{year}{1993}).

\bibitem[{\citenamefont{Arfken and Weber}(1999)}]{arfken1999mathematical}
\bibinfo{author}{\bibfnamefont{G.~B.} \bibnamefont{Arfken}} \bibnamefont{and}
  \bibinfo{author}{\bibfnamefont{H.~J.} \bibnamefont{Weber}},
  \emph{\bibinfo{title}{Mathematical methods for physicists}}
  (\bibinfo{publisher}{AAPT}, \bibinfo{year}{1999}).

\end{thebibliography}
\end{document}